\documentclass{nature}

\usepackage{scicite}
\usepackage{amsmath}
\usepackage{times}
\usepackage{txfonts}
\usepackage{color}
\usepackage{graphicx}

\title{Observation of Berry curvature in non-Hermitian system from far-field radiation}

\author
{Xuefan Yin$^{1,2\dag}$, Ye Chen$^{2\dag}$, Xiaoyu Zhang$^{2}$, Zixuan Zhang$^{2}$, Susumu Noda$^{1}$, Chao Peng$^{2,3\ast}$}

\begin{document} 



\maketitle 
\begin{enumerate}
\item Department of Electronic Science and Engineering, Kyoto University, Kyoto-Daigaku-Katsura, Nishikyo-ku, Kyoto 615-8510, Japan
\item State Key Laboratory of Advanced Optical Communication Systems and Networks, School of Electronics, $\&$ Frontiers Science Center for Nano-optoelectronics, Peking University, Beijing, 100871, China
\item Peng Cheng Laboratory, Shenzhen 518055, China
\end{enumerate}
\normalsize{$^\dag$These authors contributed equally to this work}\\
\normalsize{$^\ast$To whom correspondence should be addressed; E-mail: pengchao@pku.edu.cn}


\date{}



\begin{abstract}
Berry curvature that describes local geometrical properties of energy bands can elucidate many fascinating phenomena in solid-state, photonic, and phononic systems, given its connection to global topological invariants such as the Chern number. Despite its significance, the observation of Berry curvature poses a substantial challenging since wavefunctions are deeply embedded within the system. Here, we theoretically propose a correspondence between the geometry of far-field radiation and the underneath band topology of non-Hermitian systems, thus providing a general method to fully capture the Berry curvature without strongly disturbing the eigenstates. We further experimentally observe the Berry curvature in a honeycomb photonic crystal slab from polarimetry measurements and quantitatively obtain the non-trivial valley Chern number. Our work reveals the feasibility of retrieving the bulk band topology from escaping photons and paves the way to exploring intriguing topological landscapes in non-Hermitian systems.


\end{abstract}

Topology, namely the mathematics of conserved properties under continuous deformations, is creating a range of new opportunities throughout matters, photonics, phononics, and other wave systems\cite{hasan2010colloquium,xiao2010berry,lu_topological_2014,khanikaev_two-dimensional_2017,ozawa_topological_2019,bergholtz2021exceptional}. To characterize the topology in physics, the Berry curvature\cite{berry1984quantal,haldane2004berry} is an essential concept that describes the gauge-invariant, local, geometric manifestation of the wavefunctions in the parameter space, which is closely related to the global topological invariants such as a variety of Chern numbers\cite{thouless1982quantized,sheng2006quantum,xiao2007valley,zhang2011spontaneous,zhang2013valley}. However, since Berry curvature belongs to the intrinsic topological property of wavefunctions, it is usually deeply bound inside the system and difficult to observe. Although using tomography can reconstruct the wavefunctions in some particular scenarios \cite{hauke2014tomography,flaschner2016experimental}, much effort has been devoted to retrieving the Berry curvature from its external consequence in physics. Examples includes Hall drift in driving optical lattice \cite{price2012mapping,jotzu2014experimental,aidelsburger2015measuring} or synthetic gauge field \cite{ozawa2014anomalous,wimmer2017experimental}, Aharonov-Bohm interference of magnetic-controlled ultracold atom\cite{abanin2013interferometric,atala2013direct,duca2015aharonov,li2016bloch}, and pseudospin \cite{bleu2018measuring,gianfrate2020measurement,ren2021nontrivial,liao2021experimental,polimeno2021tuning,lempicka2022electrically} or dichroism \cite{wu2013electrical,cho2018experimental} in exciton-polariton-correlated material. Even though the specific physics varies, the mentioned observations of Berry curvature generally rely on the strong light-matter interaction to imprint the topological features of bulk wavefunction to external observables, and thus they should be categorized into the class of ``strong measurement"\cite{vallone2016strong} that the observation strongly interferes with the system. In comparison, the method of measuring the Berry curvature without much disturbing the eigenstates\cite{dressel2014colloquium} remains absent. Recall the fact that non-Hermitian photonic systems\cite{feng2017non,leykam2017edge,el2018non,shen2018topological,bergholtz2021exceptional} necessarily lose photons. The escaping photons, namely the far-field radiation, naturally carry information about the wavefunctions, thus allowing direct access to the intrinsic bulk topology that would be conventionally thought impossible. The escaping photons simply act as the ``messengers" that weakly interact with the system, but they could bridge the band topology and radiation topology to enable direct observation of Berry curvature from the far field. In recent years, radiation geometry\cite{zhen_topological_2014,alu_experimental_2018,shilei_observation_2018,chen2019singularities,yin2020manipulating,liu2021topological,shilei_polarization_singularity_2021} that concerns the non-trivial geometric structures of far-field polarization has attracted much attention because they can give rise to interesting physical consequences such as polarization half-charge around paired exceptional point\cite{zhou_observation_2018,liuwei_global_charge}, vortex beam\cite{huang2020ultrafast}, chiral devices with circular dichroism \cite{zhang2022chiral,chen2023observation}, bound states in the continuum (BICs)\cite{von_neuman_uber_1929,1985Interfering,hsu2013observation,yang2014analytical,hsu_bound_2016,jin_topologically_2019,sadreev2021interference,kang2021merging,hu2022global} and unidirectional guided resonances (UGRs)\cite{yin_observation_2020,zeng_dynamic_2021,yin2023topological}. However, whether the geometric features in the radiation originated from the band topology and how to retrieve the Berry curvature from the far field radiation still remain as elusive questions.

Here we theoretically establish a correspondence between the band topology and radiation geometry to experimentally observe the Berry curvature by characterizing the escaping photons from a non-Hermitian photonic crystal (PhC) slab system. Specifically, we prove that a full tomography of the Berry curvature can be realized by simply measuring a number of radiation channels, while for a two-level system, only one radiation channel is sufficient. Accordingly, we experimentally observe the nontrivial Berry curvatures originated from diabolic points (DPs) in a honeycomb latticed PhC slab by using a polarimetry measurement\cite{Mcmaster1954Polarization}. The Berry phases\cite{simon1983holonomy,berry1984quantal,xiao2010berry} of $\gamma \sim\pm \pi$ are obtained in an individual valley by employing integrals of the observed Berry curvature, showing nontrivial valley Chern numbers\cite{xiao2007valley,zhang2013valley} of $C_v^{\mathcal{K}} \sim \pm1/2$ as expected, and thus quantitatively validates our method. The theory and measurement also clarify that the band topology only manifests on the ``left-right" curvature\cite{shen2018topological} upon a bi-orthogonal basis of a non-Hermitian system, while a ``right-right" curvature \cite{shen2018topological} represents the geometry of radiation itself and connects to Pancharatnam-Berry (PB) phase\cite{pancharatnam1956generalized,berry1987adiabatic,lee2017recent,xie2021generalized} of far-field polarization.

{\bf The bulk-radiation correspondence of Berry curvature |} We start from a PhC slab operating in the radiation continuum as schematically illustrated in Fig. 1A, in which the $n$th photonic eigenstate $|\psi_n\rangle$ of a Hamiltonian $\hat H$ resides in the continuum. The eigenstate would radiate towards some specific directions owing to the diffraction of periodically modulated permittivity, and each gives rise to a radiation vector of $|\Psi_n\rangle$ in the far field. In a given diffraction direction, the radiation vector $|\Psi_n\rangle$ can be described by the polarization vector field of $|\Psi_n\rangle=[c_{x;n},~c_{y;n}]^T$, where $c_{x,y;n}$ are complex-valued electric-field components in $x$ and $y$ directions, respectively. \textcolor{black}{Components in $s$ and $p$ directions are discussed in Supplementary materials \cite{supp}.} Consequently, the radiation process can be understood as a linear mapping of $\mathcal{P}: |\psi_n\rangle\mapsto|\Psi_n\rangle=\hat{P}|\psi_n\rangle$, showing a direct connection between the bulk wavefunction and its radiation far-field, governed by a projection matrix denoted as $\hat{P}$.

It is well known that the Berry curvature of bulk bands can be calculated from wavefunctions as  $B_{n}=i\nabla\times\langle\phi_n|\nabla\psi_n\rangle$ (bottom panel, Fig.~1B), with $|\phi_n\rangle$ representing the left vector of $|\psi_n\rangle$.
Since our system is intrinsically non-Hermitian due to the existence of radiation losses, $|\psi_n\rangle$ and $|\phi_n\rangle$ should be bi-orthogonal to each other. \textcolor{black}{If $\hat{P}$ is an invertible matrix, we found that $|\Psi_n\rangle$ and the accompanied left vector $|\Phi_n\rangle=(\hat{P}^{-1})^\dag |\phi_n\rangle$} also form a bi-orthogonal basis for the non-Hermitian Hamiltonian of $\hat{H}^r=\hat{P}\hat{H}\hat{P}^{-1}$, leading to another Berry curvature of $B_{n}^r=i\nabla\times\langle\Phi_n|\nabla\Psi_n\rangle$ defined in the radiation field (top panel, Fig.~1B). We further prove that\cite{supp}, in the case that the matrix $\hat{P}$ is smooth ($\nabla \hat P \sim 0$) and doesn't give rise to extra vortexes \textcolor{black}{such as BICs} that carry vanished amplitudes  $\langle\Psi_n|\Psi_n\rangle =0$, the system follows  a simple correspondence between the band topology and radiation topology, given by:
\begin{eqnarray}
    \label{eq:1}  
    B_{n}^r\approx B_n
\end{eqnarray}
The above equation reveals that the escaping photons act as ``messengers" that project the bulk Berry curvature onto the far field through the matrix $\hat{P}$ (mid panel, Fig.~1B). Although the wavefunctions $|\psi_n\rangle$ and $|\phi_n\rangle$ are difficult to access because they belong to the near-field features of the eigenstates, the radiation vectors $|\Psi_n\rangle$ and $|\Phi_n\rangle$ are directly observable and can be characterized by standard optical measurement. 

In theory, the radiation in a particular diffraction direction gives a perspective projection of the wavefunctions, \textcolor{black}{and} thus it is noteworthy to discuss whether the projection is complete. For a general system that has $N$ internal degrees of freedom (DOFs), $\hat{P}$ is in the form of $2\times N$ matrix. Since one diffraction direction can only characterize two DOFs ($[c_{x;n},c_{y;n}]^T$), we probably need to measure multiple radiation channels simultaneously ---  observe the same object from different views, to fully capture the information about the bulk wavefunctions\cite{supp}. As a specific case, if the system is a simple two-level one with $N=2$,  the measurement of only one radiation channel can provide sufficient DOFs to make the projection of wavefunctions complete. Namely, we can reverse the projection process to directly determine the bulk wavefunctions from far-field radiations if $\hat{P}$ is a non-singular $2\times 2$ matrix.

To elaborate the correspondence, we consider a two-dimensional (2D) PhC slab of  Si$_3$N$_4$ material with circular air hole patterns on a honeycomb lattice (Fig. 2A). The lattice constant and slab thickness are denoted as $a$ and $h$, respectively, which give a reciprocal lattice as shown in Fig. 2B. \textcolor{black}{The grey shading area denotes the first Brillouin zone (BZ).} According to the Bloch theorem, the bulk wavefunctions can be depicted by a superposition of a series of quasi-plane-waves with discrete momentum, represented by the dots in the reciprocal lattice and we refer them as ``diffraction orders"\cite{kogelnik1972coupled}. Around the second $\mathcal{K}$ point that resides in the continuum, several diffraction orders fall into the light cone and thus open radiation channels. We take the $\mathcal{K}_1$ point of $(-4\pi/3a,0)$ as a specific example, and there exist three radiation channels $C_{1-3}$ that cause the non-Hermiticity  (red arrows, Fig. 2B). Accordingly, their in-plane momentum at $\mathcal{K}_1$ point are $\beta_1=(\sqrt{3}\beta_0/3\hat{x},0)$, $\beta_2=(-\sqrt{3}\beta_0/6\hat{x},-\beta_0/2\hat{y})$ and $\beta_3=(-\sqrt{3}\beta_0/6\hat{x},\beta_0/2\hat{y})$, respectively, with denoting $\beta_0=4\pi/\sqrt{3}a$. Three transverse-electric (TE) polarized modes can be found around the $\mathcal{K}_1$ point, and we mark them as TE$_{A, B, C}$. We assume the adjacent air holes have different radii of $r_1$ and $r_2$.  For $\delta_r=r_2-r_1=0$ that preserves the $C_{6v}$ symmetry (left panel, Fig. 2C),  it would result in a two-fold degeneracy of TE$_A$ and TE$_B$ right at the $\mathcal{K}_1$ point. As a contrary, if $r_1\neq r_2$ (right panel, Fig. 2D), the $C_{6v}$ symmetry degrades into the $C_{3v}$ symmetry that lifts the degeneracy. In the case that the in-plane symmetry breaking is sufficiently weak that allows the coupling from the TE$_C$ mode to be neglectable, the TE$_{A}$ and TE$_{B}$ modes form a two-level system \textcolor{black}{near the $\mathcal{K}_1$ point} that is described by the Hamiltonian of:
\begin{eqnarray}
\label{eq:2}
\hat{H}=\omega+\delta\sigma_z+\eta k_x\sigma_x+\eta k_y\sigma
\end{eqnarray}
in which $\omega$ is the degenerate frequency at $\mathcal{K}_1$ point; $\delta$ is related to the in-plane asymmetry and $\eta$ is the group velocity; $k_x$ and $k_y$ are dimensionless numbers to describe the momentum deviation from the $\mathcal{K}_1$ point of $\mathbf{k}=k_x\beta_0\hat{x}+k_y\beta_0\hat{y}$. Here we only consider the radiation loss but omit the material dissipation or gain, so the non-Hermiticity of $\hat{H}$ is represented by the complex degenerate frequency $\omega=\omega_r+i\gamma_0$, where $\gamma_0$ is the radiation decay rate. When $C_{6v}$ symmetry is preserved ($\delta=0$), the eigenvectors can be derived as $|\psi_n\rangle=[1, \pm |\eta| e^{i\theta}/\eta]^T$ where $e^{i\theta}=(k_x+ik_y)/|\mathbf{k}|$, creating a diabolic point at the $\mathcal{K}_1$ point which is exactly a non-Hermitian counterpart of the Dirac point in Hermitian case, and we still denote it as the DP for short without any confusion.

At the DP, the far-field polarizations of TE$_{A, B}$ bands are ill-defined since they can be mixed in arbitrary weights. Correspondingly, polarization vortexes can be found in the momentum space that each carries a half-integer topological charge (Fig. 2E), showing as a geometric feature of the DP in the far-field radiation. Once the symmetry-breaking lifts the degeneracy, the polarization vortexes degrade to a meron and anti-meron configuration\cite{guo2020meron} with circular-polarization (CPs) in opposite helicities around the $\mathcal{K}_1$ point (Fig. 2F) for TE$_{A,B}$ modes, respectively, which can be an observable signature to validate the theory. By taking the degeneracy lifting into account, the close form of \textcolor{black}{theoretical} Berry curvature in such a two-level system follows:
\begin{eqnarray}
    \label{eq:3}
    B_{n;t}=\pm \frac{4\delta\eta^2}{(4\delta^2+4\eta^2|\mathbf{k}|^2)^{3/2}}
\end{eqnarray}
\textcolor{black}{where the subscript ``$t$" distinguishes $B_{n;t}$ from the $B_n$ while the later one takes TE$_C$ into account; the signs ``$\pm$" denote the two bands $n=A,B$, respectively. Since TE$_C$ mode can be neglected in our case, we have $B_n\approx B_{n;t}$.} For a $C_6$ symmetric system ($\delta=0$), the Berry curvature shows as a $\delta$-function peaked at the $\mathcal{K}_1$ point. While for $\delta\neq 0$, the DP splits and opens a topologically nontrivial bandgap, leading to a nontrivial Berry curvature in the vicinity of $\mathcal{K}_1$ point that gives rise to non-zero valley Chern number. As we stated, such a nontrivial Berry curvature can be directly observed from the far-field radiation by employing the bulk-radiation correspondence of Eq. \ref{eq:1}. Specifically, the radiation vector $|\Psi_n\rangle$ exactly corresponds to the far-field polarization: $\Vec{S}_n(k_x,k_y)=[s_1, s_2, s_3]^T/s_0=\langle\Psi_n|\hat{\mathbf{\sigma}}|\Psi_n\rangle/s_0$, where  $\hat{\mathbf{\sigma}}=[\hat{\sigma}_z,\hat{\sigma}_x,\hat{\sigma}_y]^T$ are the Pauli's matrices and $\Vec{S}_n$ refers to the Stokes' vector in Poincare sphere which can be measured by using standard polarimetry method. Besides, the left radiation vector $|\Phi_n\rangle$ of $|\Psi_n\rangle$ can be determined from the bi-orthogonal normalization relation of $\langle\Phi_m|\Psi_n\rangle=\delta_{mn}$. As a result, we can observe the intrinsic band Berry curvature $B_{n;t}$ directly from measuring  $B_n^r$.

{\bf Experimental observation of Berry curvature |} To experimentally observe the Berry curvature, we first fabricate the PhC sample by using e-beam lithography (EBL) and inductively coupled plasma etching (ICP) processes on a Si$_3$N$_4$ slab of thickness $h=180$ nm on silica substrate (see Methods section for details). The air holes are arranged as a honeycomb lattice of $a=440$ nm with two slightly different hole radii of $r_1=50$ and $r_2=54$ nm, respectively, as the scanning electron microscope (SEM) images shown in Fig. 2A. The angle-resolved measurement system is schematically illustrated in Fig. 3A, in which a supercontinuum white light source is first sent through an acoustic-optic tunable filter (AOTF) and then linearly polarized by POL1 to generate incoherent light in a wavelength range from $550$ nm to $580$ nm. After passing through a quarter-wave plate (QWP1), the light is focused by a lens (L1) onto the rear focal plane (RFP) of an infinity-corrected objective lens ($NA=0.95$), and then illuminates the sample to excite the optical modes. \textcolor{black}{POL1 and QWP1 are used for adjusting the incident polarization for better excitation.} The radiations from the PhC sample are collected by the same objective lens and imaged by a charge-coupled device (CCD) camera which is co-focused with the RFP of the objective lens. By inserting another polarizer (POL2) and another quarter-wave plate (QWP2) before the CCD, we can fully characterize the Stokes' vector of radiation through a polarimetry method.

As shown in Fig. 3B, we found three scattered beams in the aperture of the objective lens, which correspond to the radiation channels $C_{1-3}$ plotted in Fig. 2B. To achieve the best excitation, we fine-tune the incident angle by moving L1 lens in the $x-y$ plane to illuminate the PhC sample through the channel $C_1$. \textcolor{black}{According to Fig. \ref{Fig2}F, POL1 and QWP1 adjust the incident polarization to be left-handed circular polarized (LCP) to excite TE$_A$ mode and right-handed circular polarized (RCP) to excite TE$_B$ mode, respectively.} When the AOTF selects a specific wavelength, the scatterings originating from fabrication disorders would make the iso-frequency contour at such a wavelength visible upon every radiation channel due to the on-resonance pumping mechanism\cite{regan2016direct,zhou_observation_2018,yin_observation_2020} (Fig. 3C). Accordingly, we observe the $C_3$ channel which is far from the direct reflected light and apply a cascaded $4f$ system to zoom-in the far-field pattern at a magnification rate of $\times 6$. Further, by recording the CCD images at particular arrangements of POL2 and QWP2 as polarimetry measurements, we can decompose the radiation into Stokes' vector. The result at the wavelength of  $\lambda_i=570.328$ nm is presented in Fig. 3D, in which the dashed line is the iso-frequency contour calculated from numerical simulation for visual guidance. By overlapping all the iso-frequency contours in the wavelength range of $559.383\sim 564.861$ nm for the TE$_A$ band and $565.203\sim 571.352$ nm for the TE$_B$ band, we obtain the polarization vector field in the momentum space for both TE$_{A, B}$ bands (Fig. 3E).  In the vicinity of the $\mathcal{K}_1$ point, a LCP and a RCP can be found on the TE$_A$ and TE$_B$ bands, respectively (red circles, Fig. 3E), which agree well with our prediction in theory as shown in Fig. 2F.

\textcolor{black}{Berry curvature $B^r_n$ defined in far-field radiation can be directly obtained from the full polarization vector field.} Here we consider four samples with radius differences of $\delta_r=0,~~4,~~7,~~10$ nm, and measure the $B_{A, B}^r$ of each sample, respectively (top panels, Fig. 4). \textcolor{black}{Accordingly, we calculate the numerical Berry curvature $B_{A,B}$ by employing the semi-analytical coupled-wave theory framework (CWT) \cite{liang_three-dimensional_2011,peng2012three,yang2014analytical} for a comparison (bottom panels, Fig.4).} \textcolor{black}{To better show the evolution of Berry curvatures, we plot the unit-cell geometry of each sample and the corresponding band structures of TE$_{A, B}$ as the insets in the top and bottom panels of Fig. 4.} Specifically, we start from a realistic sample with $\delta_r=0$ (Fig. 4A), where the fabrication imperfections would inevitably break the DP degeneracy at $\mathcal{K}_1$ point and give rise to a very small bandgap. We estimate that the bandgap is equivalent to the case of $\delta_r=2$ nm. In this case, we found that the $B_{A, B}^r$ show as bright spots centered at $\mathcal{K}_1$ point --- quite like $\delta$-functions (top panels, Fig.~4A), agreeing well with the numerical result $B_{A,B}$ (bottom panels, Fig.~4A). \textcolor{black}{Note that the signs of the $B_{A}^r$ and $B_{B}^r$ are exactly opposite, agreeing well with the theoretical prediction in Eq. \ref{eq:3}.} Further, we gradually open the bandgap by increasing $\delta_r$ from $4$ nm to $10$ nm (Fig.~4B). During this process, the Berry curvatures $B_{B}^r$ and $B_{B}$ gradually diffuse to a larger region in the momentum space while their peak absolute values decrease. At $\delta_r=10$ nm, the bandgap becomes quite large, and thus both $B_n^r$ and $B_n$ become fully dispersed and no longer congregate around $\mathcal{K}_1$ point. $B_{A}^r$ and $B_A$ also match well with each other and show the similar behavior (see Methods section). \textcolor{black}{The great agreement between the observed Berry curvatures $B_{A,B}^r$ and the numerical Berry curvature $B_{A,B}$ validates the bulk-radiation correspondence of band topology we propose in Eq. \ref{eq:2}.} Noteworthy that Berry curvatures are generally complex values in a non-Hermitian system. For our PhC slab in which only radiation contributes to the non-Hermiticity, the imaginary parts of Berry curvature are quite small compared to their real parts\cite{supp}.

To quantitatively validate the correspondence of Berry curvature, we calculate the geometric phases (Berry phases) $\gamma_{n}^r$ and $\gamma_n$ by applying 2D integrals over the measured and \textcolor{black}{numerical} Berry curvatures of $B_{n}^r$ and $B_{n}$ in Fig. 4, respectively. As a reference, we also derive a close-form of the \textcolor{black}{theoretical} geometric phase $\gamma_{n;t}$ in the two-level model without the contribution of TE$_C$ according to Eq. \ref{eq:3} as
\begin{eqnarray}
    \label{eq:4}
    \gamma_{n;t}=\pm\left(\pi-\frac{2\delta}{\sqrt{4\delta^2+4\eta^2k_s^2}}\pi\right)
\end{eqnarray}

\textcolor{black}{According to the valleytronics\cite{xiao2007valley}, the integral upon the individual valley region can determine the valley-Chern number (blue shading, Fig. 5A).} \textcolor{black}{Considering that the nontrivial Berry curvatures congregate around the $\mathcal{K}_1$ point, here we choose a circular integral region with radius of $k_s=0.03$ for simplicity and thus calculate $\gamma_{n}^r$ (circle), $\gamma_n$ (triangle), and $\gamma_{n;t}$ (solid line) shown in Fig. 5B and C. According to Eq.\ref{eq:4}, the close-form geometric phases $\gamma_{n;t}$ exactly equals to $\pm\pi$ at $\delta_r=0$ for TE$_{A, B}$ bands, respectively\cite{zhang2005experimental}, corresponding to the nontrivial, quantized, valley-Chern number of $C_v^{\mathcal{K}}=\pm 1/2$ in an individual valley\cite{xiao2007valley}. When $\delta_r\neq 0$, the open bandgap (nonzero $\delta$) would make the geometric phases deviate from $\pm\pi$, unless the integral region $k_s$ tends to be infinite\cite{xiao2010berry}. Such a behavior can be verified by our experimental observation. Specifically, the three geometric phases $\gamma_{n}^r$, $\gamma_{n}$ and $\gamma_{n;t}$ of both TE$_{A, B}$ bands agree well with each other, quantitatively confirming the validity of bulk-radiation correspondence we propose in Eq. \ref{eq:2}. We also found $\gamma_{A,B}^r \approx \gamma_{A,B} \approx \gamma_{A,B;t}\approx \pm 0.6\pi$ at $\delta_r=10$ nm, clearly showing the impact of nonzero bandgap.} Besides, when $\delta_r$ is considerable large, we notice that the theoretical phase $\gamma_{n;t}$ from the two-level model slightly deviates from the numerical phase $\gamma_{n}$. This is because the influence of the TE$_C$ band is neglected in the derivation of $\gamma_{n;t}$. The discussion on measuring Berry curvature in a system with higher DOFs ($N>2$) is presented in Supplementary materials\cite{supp}.

In the experiment stated above, we obtain the left vector $|\Phi_n\rangle$ by using the bi-orthogonal normalization relation of $\langle\Phi_n | \Psi_n\rangle=1$. However, the left vector can also be directly measured. According to the reciprocity, the left vector at the $\mathbf{k}$ point corresponds to the right vector at the $-\mathbf{k}$ point, namely $|\Phi_n(\mathbf{k})\rangle=|\Psi^{*}_n(-\mathbf{k})\rangle$\cite{supp}. Therefore, we can also observe one band at both the $\mathbf{k}$ and $-\mathbf{k}$ points to retrieve the system's topology, instead of measuring two bands simultaneously. We also emphasize that the radiation Berry curvature $B_n^r$ directly corresponds to the bulk topology of $B_n$ only when the projection matrix $\hat{P}$ doesn't cause extra geometric phases, which is the case of our experiment (\textcolor{black}{$\nabla\hat{P}\approx 0$ around} the $\mathcal{K}$ point).

We note that other than the aforementioned ``left-right" curvature \textcolor{black}{$B_n^r$ which depicts the bulk band topology}, we can also define a ``right-right" curvature of $B_{n;rr}^r=i\nabla\times\langle\Psi_n|\nabla\Psi_n\rangle$ to capture the geometric features of the polarization field itself. The integral over ``right-right" curvature directly presents the PB phase of far-field polarization, showing the swirling structure of Stokes' vector $\Vec{S}_n$. For instance, as shown in Fig. 3E, we can find meron and anti-meron features that are originated from the DP.  Such nontrivial polarization features don't directly correspond to the valley-Chern number defined on bulk Berry curvature (``left-right" curvature). Instead, they are related to the Skyrmion number\cite{skyrme1961particle,skyrme1962unified,fert2017magnetic} that is associated with the PB phase and ``right-right" curvature (See Methods section for details).

{\bf Conclusion |} In summary, our findings of the ``bulk-radiation correspondence" of Berry curvature reveal the feasibility of retrieving the band topology from characterizing escaping photons in far-field radiation. We prove in theory and demonstrate in experiments that \textcolor{black}{characterizing the polarizations of }one radiation channel can capture a complete map of the eigenstates in a two-level non-Hermitian system, to directly access the Berry curvature and Chern number without strongly disturbing the system. The proposed method can also be extended to multi-level systems by measuring more radiation channels simultaneously\cite{supp}, and utilized to extract other topological features such as quantum geometric tensor\cite{provost1980riemannian,anandan1990geometry,bleu2018measuring,gianfrate2020measurement,liao2021experimental,ren2021nontrivial}. Our work demonstrates a simple and effective way of directly observing the Berry curvature in non-Hermitian systems and thus could shed light on the exploration of the intriguing phases in topological systems.

\bibliography{scibib}

\bibliographystyle{naturemag}

\clearpage
\begin{figure}[htbp] 
 \centering 
 \includegraphics[width=16cm]{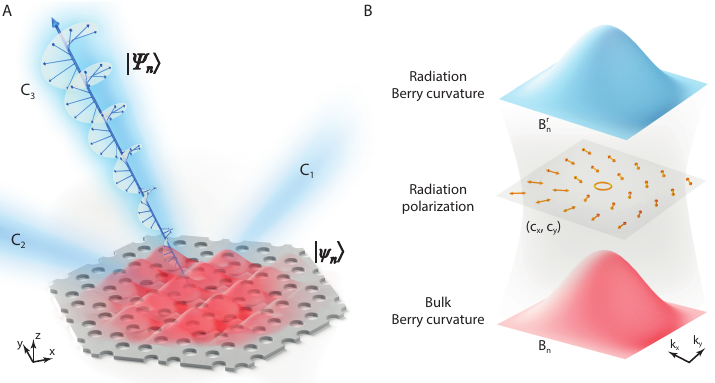} 
\caption{
\textbf{Correspondence between bulk band topology and far-field radiation.} (A) The schematic of radiation process from PhC slab to the far field in real space. The wavefunction of the optical eigenmodes $|\psi_n\rangle$ in PhC slab is diffracted by the periodic lattice into several specific diffraction directions $C_{1-3}$, acting as the radiation channels. For one channel (i.e. $C_3$), the radiation vector of $|\Psi_n\rangle$ can be defined in the polarization of the diffracted wave, marked as the spiral arrows. (B) \textcolor{black}{The ``bulk-radiation correspondence" of Berry curvature in momentum space.} The radiation polarization field (middle panel) can bridge the Berry curvature $B_n$ defined in wavefunction $|\psi_n\rangle$ (bottom panel) with the Berry curvature $B_n^r$ defined in far-field radiation vector $|\Psi_n\rangle$ (top panel). $c_{x,y}$ are complex amplitudes of the radiative waves in $x-y$ plane.}
\label{Fig1}
\end{figure}

\clearpage
\begin{figure}[htbp] 
 \centering 
 \includegraphics[width=16cm]{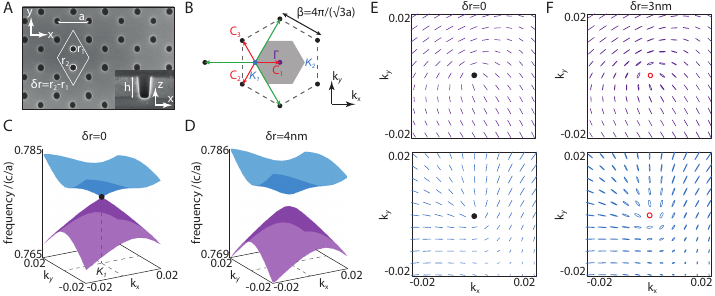} 
\caption{\textbf{Demonstration of \textcolor{black}{Berry curvature observation} on a honeycomb latticed PhC slab.} (A) SEM image of the fabricated PhC sample, showing a honeycomb latticed structure of SiN slab on SiO$_2$ substrate with different hole radii. Inset: the side view of the air hole. The structural parameters are given as $a=440$ nm, $r_1=50$ nm, $r_2=54$ nm, $h=180$ nm, respectively. $\delta_r$ is defined as $r_2-r_1$. (B) The reciprocal lattice of the PhC sample. Grey shading area: the first BZ; purple dot: the $\Gamma$ point; blue dot: the $\mathcal{K}_1$ point; red vectors: three diffraction orders acting as radiation channels $C_{1-3}$; green vectors: non-radiative basic diffraction orders. (C, D) The band structures of PhCs around $\mathcal{K}_1$ point with ($\delta_r=0$) and without ($\delta_r=4$ nm) the inversion symmetry. Owing to the $C_6$ symmetry with $\delta_r=0$, TE$_A$ mode (purple sheet) and TE$_B$ mode (blue sheet) are degenerate at $\mathcal{K}_1$ point, giving rise to a diabolic point. When $\delta_r\neq 0$, the $C_6$ symmetry degrades to $C_3$ symmetry, and thus the DP split to open a non-trivial band gap between TE$_{A,B}$ modes. (E, F) The polarization fields in the momentum space of TE$_A$ (purple, the top panels) and TE$_B$ (blue, the bottom panels) modes around $\mathcal{K}_1$ point with ($\delta_r=0$) or without ($\delta_r=4$nm) the inversion symmetry, respectively. For $\delta_r=0$ that maintains the $C_6$ symmetry, half charges emerge at $\mathcal{K}_1$ point due to the DP for both TE$_{A,B}$ modes. When $\delta_r\neq 0$, two CPs with opposite handness emerge around the $\mathcal{K}_1$ point instead. Black dot: DP; red marks: the \textcolor{black}{quasi-CPs}. All data are calculated by numerical simulations (COMSOL Multiphysics).}
\label{Fig2}
\end{figure}

\clearpage
\begin{figure}[htbp] 
 \centering 
 \includegraphics[width=16cm]{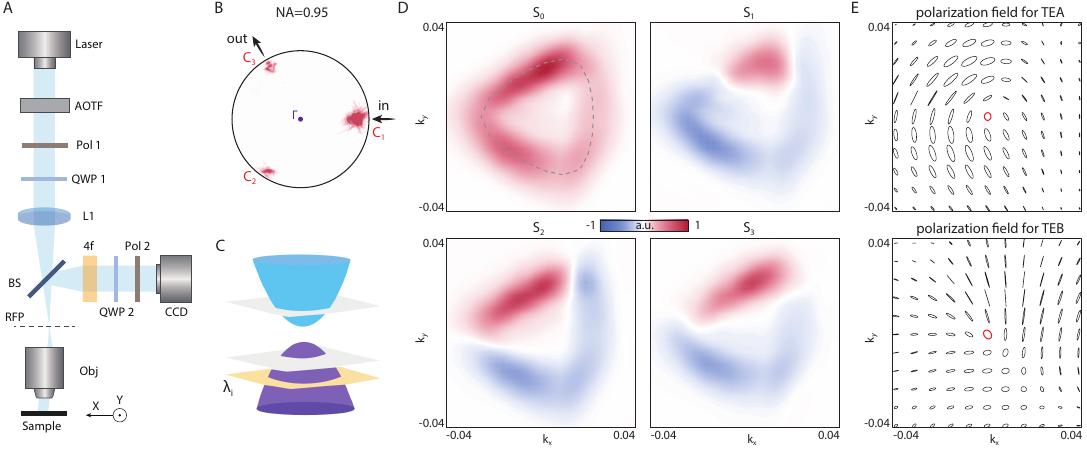} 
\caption{\textbf{Polarimetry measurement of far-field polarization fields.} (A) Schematic of the measurements setup. AOTF: acoustic-optic tunable filter; POL1 and POL2: polarizers; QWP1 and QWP2: quarter-wave plates; L1: convex lens with $20$ cm focal length; BS: beam splitter; RFP: rear focal plane; Obj: Objective lens with NA of 0.95 and working distance of 150 $\mu$m; $4f$: lens system with magnification rate of $\times$6. (B) The observed image of three scattered beams from radiation channels $C_{1-3}$ within the NA range. In the experiment, we excite the optical modes from channel $C_1$ by moving lens L1 to a proper position and then collect the diffracted lights in channel $C_3$ after it is magnified by the $4f$ system. (C) Schematic of isofrequency contours of the two-level system near the $\mathcal{K}_1$ point. Yellow plane denotes the wavelength of $\lambda_i=570.328$ nm. Purple sheet: TE$_A$ mode; blue sheet: TE$_B$ mode. (D) The measured isofrequency contour $S_0$ and Stokes' parameters $S_{1-4}$ at wavelength of $\lambda_i=570.328$ nm. The Stokes' parameters are determined through different configurations of POL2 and QWP2. Dashed line: the simulated iso-frequency contour at $\lambda_i$. (E) The measured polarization distributions in the momentum space around the $\mathcal{K}_1$ point, by overlapping several isofrequency contours and evaluating the overall Stokes' parameters. Red marks: the quasi-CPs.}
\label{Fig3}
\end{figure}

\clearpage
\begin{figure}[htbp] 
 \centering 
 \includegraphics[width=16cm]{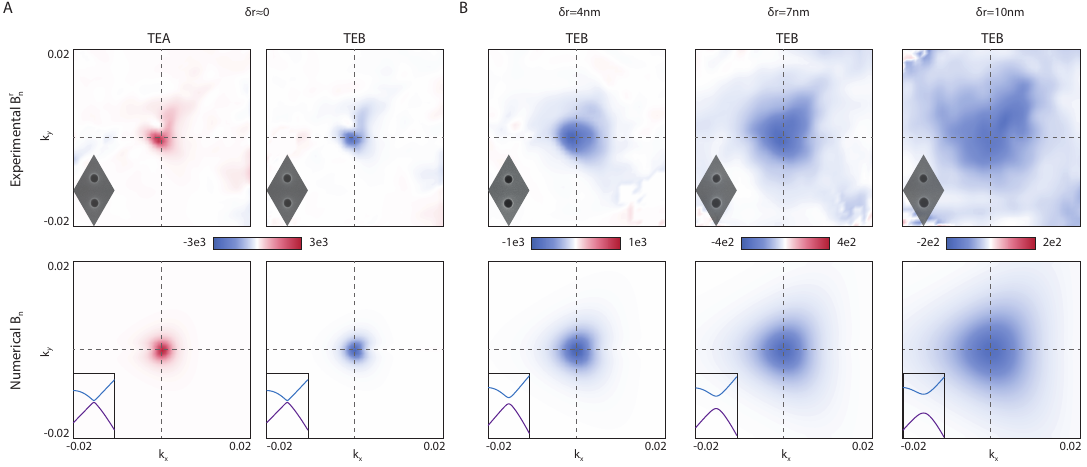} 
\caption{\textbf{Experimental observation of Berry curvatures.} (A) The measured Berry curvatures $B_n^r$ from far-field radiation for the PhC sample with $\delta_r\approx 0$ (top panels) and the numerically calculated (semi-analytical CWT) Berry curvatures $B_n$  with $\delta_r=2$ nm (bottom panels) for a comparison. For a realistic sample with $\delta_r=0$, the fabrication errors would slightly lift the DP at $\mathcal{K}_1$ point to create a small bandgap. We estimate that the bandgap is equivalent to the case of $\delta_r=2$ nm. In this case, the Berry curvatures congregate around the $\mathcal{K}_1$ point since the bandgap is very small, showing opposite signs for TE$_A$ and TE$_B$ modes. (B) The measured Berry curvatures $B_B^r$ from far-field radiation (top panels) and the numerically calculated bulk Berry curvatures $B_B$ (bottom panels) for TE$_B$ mode when $\delta_r=4$ nm (left), $7$ nm (middle) and $10$ nm (right). Along with the increasing $\delta_r$, the bandgap gradually opens and the Berry curvature gradually diffuses to a larger region in the momentum space. Insets in top panels: SEM images of unit cell of each PhC sample; insets in bottom panels: band structures of the two-level system accordingly.}
\label{Fig4}
\end{figure}

\clearpage
\begin{figure}[htbp] 
 \centering 
 \includegraphics[width=16cm]{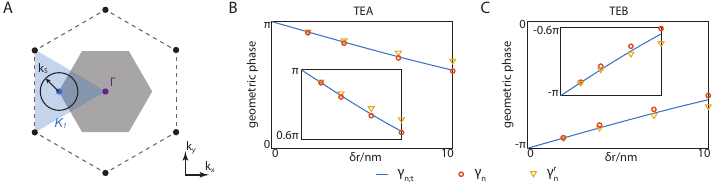} 
\caption{\textbf{Geometric phases obtained from measured Berry curvatures.} (A) Schematic of an individual valley (blue shading) around the $\mathcal{K}_1$ point (blue dot) in the reciprocal lattice. The integral on Berry curvature over the individual valley gives the geometric (Berry) phases. Considering that the Berry curvatures congregate around the $\mathcal{K}_1$ point when $\delta_r$ is relatively small, we perform the integral on a circular region (black circle) to simplify the calculation. Grey shading: the first BZ; purple dot: the $\Gamma$ point. (B, C) The geometric phases for TE$_{A,B}$ modes. Blue solid line: theoretical Berry phase $\gamma_{n;t}$ of the two-level model according to Eq. \ref{eq:4}; red circles: numerical Berry phase $\gamma_n$ obtained from the integral of Berry curvatures $B_n$ that are calculated by the CWT; yellow triangles: geometric phase $\gamma_{n}^r$ obtained from measured Berry curvature $B_{n}^r$ shown in Fig. 4. When $\delta_r=0$, the Berry phases are exactly $\pm\pi$ for TE$_{A,B}$ modes owing to the existence of DP, corresponding to quantized valley-Chern numbers of $\pm 1/2$. When $\delta_r$ gradually increases, the calculated and measured Berry phases both gradually deviate from the quantized $\pm\pi$. Moreover, when $\delta_r$ becomes relatively large, the theoretical $B_{n;t}$ slightly deviates from the numerical $B_n$, due to the impact of TE$_C$ mode.}
\label{Fig5}
\end{figure}

\end{document}